\documentclass[aps,twocolumn,showpacs,prl,10pt,superscriptaddress,preprintnumbers,nofootinbib]{revtex4-2}
\usepackage{amsmath, amssymb}
\usepackage{physics, bm}
\usepackage{graphicx}
\usepackage{xcolor}
\usepackage[colorlinks, linkcolor=red, citecolor=red, urlcolor=magenta]{hyperref}

\usepackage[normalem]{ulem}
\usepackage{soul}
\usepackage{multirow}
\usepackage{orcidlink}

\newcommand{\nn}{\nonumber}
\begin{document}
%%%%%%%%%%%%%%%%%%%%%%%%%%%%%%%%%%%%%%%%%%%%%%%%%%%%%%%%%%%%%%%%

\preprint{
	{\vbox {
		\hbox{\bf CPTNP-2025-045}
}}}
\vspace*{0.2cm}

%================================================================================================
\title{Unveiling Light-Quark Yukawa Flavor  Structure via Dihadron Fragmentation at Lepton Colliders}
%================================================================================================

\author{Qing-Hong Cao\,\orcidlink{0000-0003-0033-2665}}
\email{qinghongcao@pku.edu.cn}
\affiliation{School of Physics, Peking University, Beijing 100871, China}
\affiliation{School of Physics, Zhengzhou University, Zhengzhou 450001, China}
\affiliation{Center for High Energy Physics, Peking University, Beijing 100871, China}

\author{Xin-Kai Wen\,\orcidlink{0009-0008-2443-5320}}
\email{xinkaiwen@ihep.ac.cn (corresponding author)}
\affiliation{Institute of High Energy Physics, Chinese Academy of Sciences, Beijing 100049, China}
\affiliation{China Center of Advanced Science and Technology, Beijing 100190, China}

\author{Bin Yan\,\orcidlink{0000-0001-7515-6649}}
\email{yanbin@ihep.ac.cn (corresponding author)}
\affiliation{Institute of High Energy Physics, Chinese Academy of Sciences, Beijing 100049, China}
\affiliation{Center for High Energy Physics, Peking University, Beijing 100871, China}

\author{Shu-Tao Zhang\,\orcidlink{0009-0009-5269-2364}}
\email{shutaozhang@pku.edu.cn}
\affiliation{School of Physics, Peking University, Beijing 100871, China}

\date{\today}
%================================================================================================
\begin{abstract}
Directly probing light-quark Yukawa couplings and their flavor structure remains a major challenge due to their smallness and overwhelming QCD backgrounds. In this Letter, we propose a theoretical framework to access these couplings at lepton colliders through transverse spin dependent azimuthal modulations in dihadron fragmentation. 
These modulations arise from the interference between Higgs mediated and standard model amplitudes in $e^-e^+\to q\bar{q}Z$, producing angular structures that are linearly sensitive to the Yukawa couplings $y_q$, in contrast to conventional observables that scale as $y_q^2$. By combining channels with an identified accompanying single hadron,  $h^\prime=\pi^\pm,K^\pm$, and $p/\bar{p}$, this approach cleanly disentangles the up- and down-quark Yukawa contributions, yielding typical limits at the $\mathcal{O}(10^{-4}\sim 10^{-3})$ level and establishing fragmentation dynamics as a novel and complementary probe of the Higgs flavor structure.
\end{abstract}
%================================================================================================

\maketitle

%================================================================================================
\emph{Introduction---}
%================================================================================================
The Higgs mechanism is a central element of the standard model (SM), generating fermion masses through Yukawa interactions and predicting the relation $y_f=\sqrt{2}m_f/v$ with $v=246~{\rm GeV}$. The Yukawa interactions of the Higgs boson with third-generation charged fermions have been firmly established at the Large Hadron Collider (LHC)~\cite{Cao:2016wib,Cao:2019ygh,Li:2019uyy,Cao:2020hhb,Bi:2020frc,ATLAS:2022vkf,CMS:2022dwd,CMS:2022kdi,ATLAS:2025eua,ATLAS:2024wfv,ATLAS:2024yzu}.  Sensitivity to second-generation couplings is also rapidly improving~\cite{Bodwin:2013gca,Kagan:2014ila,Konig:2015qat,Brivio:2015fxa,ATLAS:2024ext,ATLAS:2024yzu,CMS:2025rxu}, and evidence for the Higgs boson decays to dimuons has recently been observed with a significance of $3.4\sigma$ using combined  Run-2  and Run-3 data~\cite{ATLAS:2025coj}. In contrast, the Yukawa couplings of first-generation fermions, in particular the light quarks, remain essentially unconstrained.  Current bounds still allow values that differ from their SM predictions by orders of magnitude, leaving a major gap in our experimental verification of the Higgs mechanism~\cite{CMS:2023rcv,CMS:2025xkn}.  Such large modifications are well motivated in many extensions of the SM~\cite{Bar-Shalom:2018rjs,Erdelyi:2024sls}, making it essential to close this gap in order to confirm the universality of the Higgs mechanism for all fermion mass generation and to probe potential new physics (NP) in the Yukawa sector.

Directly measuring the light-quark Yukawa couplings, however, is extremely challenging, owing to the overwhelming QCD backgrounds in hadronic Higgs decays and the tiny sizes of the couplings. Consequently, most existing studies rely on indirect constraints derived from precision measurements of Higgs production rates, kinematic distributions, rare decays, and global analyses within the effective field theory framework~\cite{Zhou:2015wra,Bishara:2016jga,Soreq:2016rae,Gao:2016jcm,Yu:2016rvv,Falkowski:2020znk,Vignaroli:2022fqh,Alasfar:2022vqw,Balzani:2023jas,Yan:2023xsd,Michel:2025afc}.  Current analyses yield limits of  $|y_u|<0.030$ and $|y_d|<0.026$, assuming that all other Higgs couplings agree with their SM values and only one light quark Yukawa coupling is varied at a time~\cite{CMS:2023rcv,CMS:2025xkn}. These bounds are expected to improve at the high-luminosity LHC, reaching sensitivities of $|y_{u,d}|<0.007$~\cite{deBlas:2019rxi}. However, such constraints do not provide independent sensitivity to the up- and down-quark Yukawa couplings because experimentally distinguishing their respective contributions in any Yukawa sensitive process is extremely challenging. As a result, the constraints effectively apply only to specific  combinations of light-quark flavors.

To address these challenges, we propose in this Letter a novel strategy for directly probing the up- and down-quark Yukawa couplings by exploiting interference between the Higgs induced signal and the SM background. Our approach leverages the transverse spin effects of light quarks generated by Yukawa interactions at lepton colliders, which depend linearly on $y_q$. This linear dependence provides a potential advantage over conventional methods, which are limited by cross section or decay rate measurements that scale as $y_q^2$.  
The transverse spin of a final-state quark can be accessed through dihadron fragmentation process under the collinear factorization framework~\cite{Collins:1993kq,Collins:2011zzd}. Recently, such effects arising from non-perturbative QCD dynamics have emerged as a novel tool for improving SM measurements and probing  NP~\cite{Zhou:2011ba,Boughezal:2023ooo,Wang:2024zns,Wen:2024cfu,Wen:2024nff,Yang:2024kjn,Cheng:2025cuv,Michel:2025afc,Huang:2025ljp,Cao:2025qua,Qin:2025cvp,He:2026pxa}.  In particular, we show that the associated production of a dihadron with an additional hadron $h^\prime$ (e.g., $h^\prime=\pi^\pm,K^\pm, p/\bar{p}$) provides an effective way to realize the interference and disentangle the up- and down-quark Yukawa couplings, thereby enhancing sensitivity and revealing the flavor structure.

\begin{figure}[htbp]
	\centering
	\includegraphics[scale=0.4]{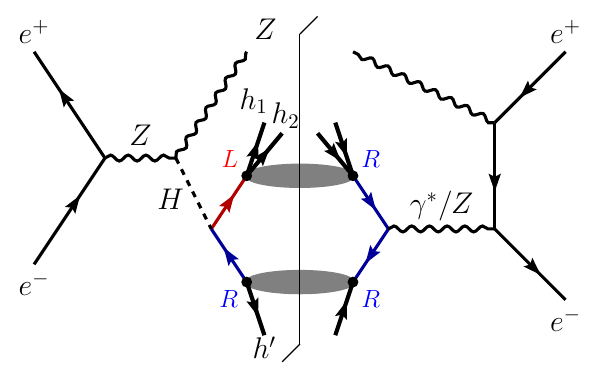}
	\includegraphics[scale=0.4]{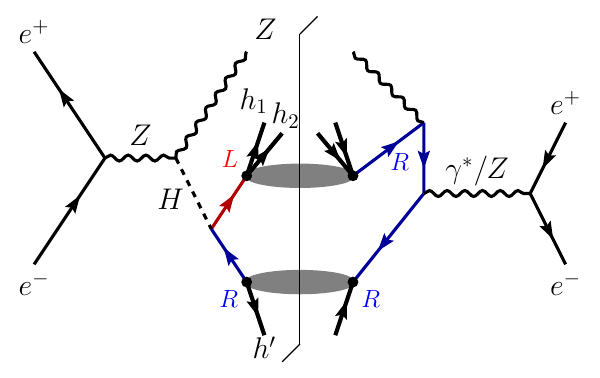}
	\caption{Representative diagrams for the interference between the Higgs signal and continuum background with identified fragmentation hadrons in Eq.~\eqref{eq:sia}. Gray bubbles indicate additional fragmentation products. Quark chiralities $(L,R)$ are shown, and the Yukawa coupling induces a chirality flip.}
	\label{Fig:MM}
\end{figure}

%================================================================================================
\vspace{3mm}
\emph{Kinematics and observable---}
%================================================================================================
We investigate the inclusive associated production of a $Z$ boson with a light-quark pair in $e^-e^+$ collisions, $e^- e^+ \to q\bar qZ$,
as a direct probe of the light-quark Yukawa couplings, which  are described by the effective Lagrangian 
\begin{equation}\label{eq:Lagrangian}
	\mathcal{L}_{\rm int} = -\frac{y_q}{\sqrt{2}} H \bar{\psi}_q \psi_q - i \frac{\tilde{y}_q}{\sqrt{2}} H \bar{\psi}_q \gamma_5 \psi_q,
\end{equation}
where $H$ denotes the Higgs field, and $y_q$ ($\tilde{y}_q$) parametrizes the $CP$-even ($CP$-odd) component. 
The Higgs-induced contribution arises from the golden process $e^+e^-\to ZH$ followed by $H\to q\bar q$, while the dominant background originates from continuum process $e^- e^+ \to q\bar q$ with initial-state and final-state radiation, as shown in Fig.~\ref{Fig:MM}. 
Their interference induces a single helicity flip of light quark, leading to a transverse-spin dependent contribution that is linear in $y_q$ and $\tilde{y}_q$.  

To isolate this transverse-spin effect and disentangle the flavor dependence of the Yukawa couplings, we consider events in which one of the final-state partons fragments into a dihadron while the recoiling parton produces a single hadron. In the center-of-mass frame (c.m.), the process is
\begin{equation}\label{eq:sia}
	e^{-}(\ell) + e^{+}(\ell^\prime)  \rightarrow  [h_1(p_1) + h_2(p_2) ] + h'(p') + Z(P_Z) + X,
\end{equation}
where the momenta of the corresponding particles are indicated in parentheses. The hadrons $h_1$, $h_2$, and $h^\prime$ carry large energies. The pair $(h_1,h_2)$ is required to be collimated and well separated from $h^\prime$, ensuring that it originates from the fragmentation of a single parton while $h^\prime$ identifies the recoiling parton. This topology provides a clean projection of the quark transverse spin, and the identity of the tagging hadron $h^\prime$ provides flavor discrimination among light-quark Yukawa couplings.

%------------------------------------------------------------------------
%------------------------------------------------------------------------

\begin{figure}
	\centering
	\includegraphics[scale=0.5]{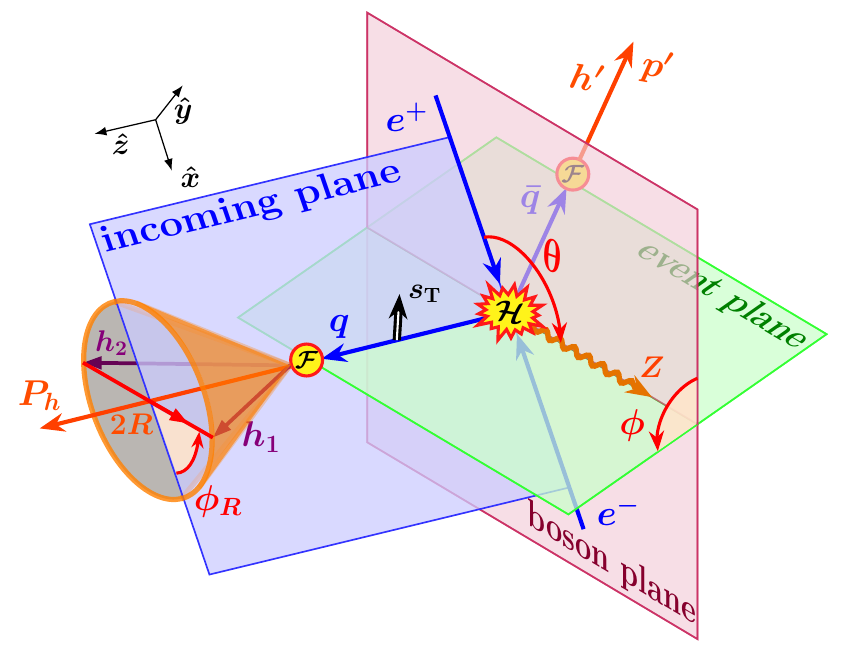}
	\caption{Kinematic configuration of $e^-e^+\to q\bar{q}Z$ with quark $q$ fragmenting into a dihadron ($h_1h_2$) and antiquark $\bar{q}$ producing a single hadron $h'$.}
	\label{Fig:Geo}
\end{figure}

To characterize the dihadron kinematics, we construct a local coordinate system based on the combined momentum $P_h = p_1+p_2$. The axes are defined as
\begin{equation}\label{eq:frame}
	\hat{z} = \frac{\bm{P}_h}{|\bm{P}_h|}, \quad
	\hat{y} = \frac{\bm{\ell} \cross \hat{z}}{|\bm{\ell} \cross \hat{z}|}, \quad
	\hat{x} = \hat{y} \cross \hat{z}.
\end{equation}
The transverse vector $\bm{R}_T$  is the projection of the relative momentum $R^\mu=(p_1^\mu-p_2^\mu)/2$ onto the $\hat{x}-\hat{y}$ plane and defines the azimuthal angle $\phi_R$, as illustrated in Fig.~\ref{Fig:Geo}. We further introduce $q_\ell=\ell+\ell^\prime$ and define the kinematic invariants,
\begin{equation*}
	s = q_{\ell}^2, \quad 
	x_q= \frac{2 P_q \cdot q_{\ell}}{s} = \frac{2 E_q }{\sqrt{s}} ,\quad
	x_{\bar q}=  \frac{2 P_{\bar q} \cdot q_{\ell}}{s} = \frac{2 E_{\bar q} }{\sqrt{s}},
\end{equation*}
where $P_q$ and $P_{\bar q}$ denote the quark and antiquark momenta, and $E_q$ and $E_{\bar q}$ are their energies in the c.m. frame. The variables $x_q$ and $x_{\bar q}$ thus represent the 
energy fractions of quarks. Similarly, the hadronic energy fractions are
\begin{equation*}
z=\frac{P_h\cdot q_\ell}{P_q\cdot q_\ell}=\frac{P_h^0}{E_q},\quad \bar{z}=\frac{p^\prime\cdot q_\ell}{P_{\bar q}\cdot q_\ell}=\frac{p^{\prime 0}}{E_{\bar q}}.
\end{equation*}
These variables can be expressed in terms of hadronic-level observables $P_h^0$, $p^{\prime 0}$, $\alpha$, and $E_Z$ in the c.m.~frame from momentum conservation, where $\alpha$ and $E_Z$ denote the angle between the $Z$ boson and the single-hadron $h^\prime$, and the $Z$-boson energy, respectively.

%------------------------------------------------------------------------
%------------------------------------------------------------------------

In the collinear factorization region, defined by the hierarchy $M_h \ll |\bm{P}_h|$, $m_{h'} \ll |\bm{p}'|$, and $M_h$, $m_{h'} \ll \sqrt{(P_h + p')^2}$,  where $M_h^2 = P_h^2$ and $m_{h'}^2=p'^2$ are the invariant masses of the dihadron $(h_1 h_2)$ and hadron $h'$, respectively~\cite{Collins:1993kq,Collins:2011zzd}, the differential distribution for the process in Eq.~\eqref{eq:sia} can be expressed as
\begin{equation}\label{eq:diXsec}
	\begin{aligned}
		&\frac{d\sigma}{dx_q \, dx_{\bar q} \, d\cos\theta \, d\phi \, dz \, d\bar{z} \, dM_h \, d\phi_R} 
		= \frac{1}{64(2\pi)^5} \sum_{q, \, q\to \bar{q}} \, \mathcal{C}_q D^{h^\prime}_{\bar{q} }(\bar{z}) \\
		&  \times \big[ D^{h_1 h_2}_{q}(z, M_h)-(\bm{s}_{T,q}\times \hat{\bm R}_T)^zH^{h_1 h_2}_{q}(z, M_h) \big].
	\end{aligned}
\end{equation}
Here, $\theta$ is the polar angle between the incoming electron and the produced $Z$ boson, and $\phi$ denotes the rotation of the event plane around the $Z$-boson axis (see Fig.~\ref{Fig:Geo}). Together with the energy fractions $x_q$ and $x_{\bar q}$, these angular variables fully determine the three-body kinematics of the hard scattering. Consequently, all final-state momenta can be reconstructed  using the spherical law of cosines and the momentum conservation~\cite{wen:long}.  For example, the  polar and azimuthal angles of the quark in the c.m. frame are given by
\begin{align}
&\cos\theta_q=\cos\theta\cos\theta_{qZ}+\sin\theta\sin\theta_{qZ}\cos\phi, \nn\\
&\frac{\sin\phi_q}{\sin\theta_{qZ}}=-\frac{\sin\phi}{\sin\theta_q}, \quad \cos\theta_{qZ}=-\frac{x_Z^2+x_q^2-x_{\bar q}^2-\beta}{2 x_q\sqrt{x_Z^2-\beta}},
\end{align}
where $x_Z=2-x_q-x_{\bar q}$ is the energy fraction of the $Z$ boson and $\beta=4m_Z^2/s$, 
while $\theta_{qZ}$ denotes the opening angle between the quark and the $Z$ boson. The corresponding angles for the antiquark ($\theta_{\bar q},\phi_{\bar{q}}$)  can be obtained by the replacements $x_q\leftrightarrow  x_{\bar q}$ and $\phi\to \pi +\phi$.

The hard coefficient $\mathcal{C}_q$ and the quark transverse spin vector $\bm{s}_{T,q}=(s_q^x,s_q^y)$ in Eq.~\eqref{eq:diXsec} are determined from the helicity amplitudes $\mathcal{M}_{\lambda\bar{\lambda}\lambda_Z}^{\lambda_1\lambda_2}$ for $e^-_{\lambda_1}e^+_{\lambda_2}\to q_\lambda\bar{q}_{\bar \lambda}Z_{\lambda_Z}$, with $\lambda_i$ denoting particle helicities. Their spin structure is encoded in the quark density matrix,
\begin{equation}
\mathcal{H}_{\lambda\lambda^\prime}^q\equiv \frac{1}{4}\mathcal{M}_{\lambda\bar{\lambda}\lambda_Z}^{\lambda_1\lambda_2}(\mathcal{M}_{\lambda^\prime\bar{\lambda}\lambda_Z}^{\lambda_1\lambda_2})^*\equiv\frac{\mathcal{C}_q}{2}(\delta_{\lambda\lambda^\prime}+s_q^i\sigma_{\lambda\lambda^\prime}^i),
\label{eq:density}
\end{equation}
where repeated indices are summed and $\sigma^i=(\sigma^x,\sigma^y,\sigma^z)$ are the Pauli matrices. A transversely polarized quark generated by the Yukawa interaction subsequently fragments into a dihadron, producing the distinctive azimuthal modulation
\begin{equation}
( \bm{s}_{T, q} \times \hat{\bm{R}}_T )^z = s^x_{q} \sin\phi_R - s^y_{q} \cos\phi_R,
\end{equation}
where $\hat{\bm{R}}_T =\bm{R}_T/|\bm{R}_T|$. The associated $\cos\phi_R$ and $\sin\phi_R$  terms thus provide clean and robust signatures of quark transverse polarization. 
At the leading order (LO), the transverse spins are given by
\begin{align}
	s^x_q \, \mathcal{C}_q  &= 2 \Re \mathcal{H}^q_{+-}
	=  \omega_q \, y_q + \tilde{\omega}_q \, \tilde{y}_q  , \nn\\
	s^y_q \, \mathcal{C}_q &= 2 \Im \mathcal{H}^q_{+-}
	=  \tilde{\omega}_q \, y_q - \omega_q \, \tilde{y}_q  ,
	\label{eq:H-sxy}
\end{align}
where $\omega_q$ and $\tilde{\omega}_q$ are real functions of the hard scattering kinematics. This structure reflects the linear dependence of $\mathcal{H}_{+-}^q$ on $y_q-i\tilde{y}_q$. In contrast to the two-body case where $s_q^x$ is purely $CP$ even and $s_q^y$ purely $CP$ odd~\cite{Yu:2023shd}, both components here receive linear contributions from $y_q$ and $\tilde{y}_q$, generating intrinsic  $CP$ mixing. Consequently, $CP$-violating effects cannot be extracted from $\phi_R$ distribution alone.
Analogous results hold for the antiquark under a  $CP$ transformation.

The hadronization process is described by fragmentation functions (FFs), which encode the probability of a parton producing specific hadrons~\cite{Berman:1971xz,Field:1977fa,Feynman:1978dt}.  For a dihadron system, $D^{h_1 h_2}_{q}(z, M_h)$ and $H^{h_1 h_2}_{q}(z, M_h)$ in Eq.~\eqref{eq:diXsec}  correspond to the  unpolarized dihadron FF (DiFF) and the interference DiFF, respectively. The function $D^{h_1h_2}_q$ determines the production rate, while the ratio $H^{h_1 h_2}_{q} / D^{h_1 h_2}_{q}$ quantifies the transverse spin analyzing power. Similarly, the fragmentation of the antiquark into a single hadron is described by the unpolarized FF $D^{h'}_{\bar{q} }(\bar{z})$.  All FFs depend on the  factorization scale $\mu$, which we take to be the Higgs mass and whose scale dependence is suppressed here for simplicity.

%================================================================================================
\vspace{3mm}
\emph{Numerical results and discussion---}
%================================================================================================
Below, we investigate the sensitivity of future lepton colliders to light-quark Yukawa couplings through azimuthal asymmetries in the process of Eq.~\eqref{eq:sia}. As a benchmark, we consider $\sqrt{s}=250~{\rm GeV}$ and an integrated luminosity of $\mathcal{L}=50~{\rm ab}^{-1}$~\cite{Ai:2025cpj}, which is representative of proposed facilities such as the Circular Electron-Position Collider (CEPC)~\cite{CEPCStudyGroup:2018ghi,CEPCStudyGroup:2023quu}, International Linear Collider (ILC)~\cite{ILC:2013jhg}, and the Future Circular Collider (FCC-ee)~\cite{FCC:2018evy}.  Currently, only the $(h_1 h_2) = (\pi^+\pi^-)$ DiFFs are available~\cite{Pitonyak:2023gjx, Cocuzza:2023oam, Cocuzza:2023vqs}, which satisfy isospin and  charge conjugate symmetries:
\begin{gather}
	D^{\pi^+\pi^-}_{u} = D^{\pi^+\pi^-}_{d}, \;\;
	H^{\pi^+\pi^-}_{u} = -H^{\pi^+\pi^-}_{d},	\;\;
	H^{\pi^+\pi^-}_{s, \bar{s}, c, \bar{c}, b, \bar{b}} = 0, \nn\\
	D^{\pi^+\pi^-}_{q} = D^{\pi^+\pi^-}_{\bar{q}}, \;\;
	H^{\pi^+\pi^-}_{q} =- H^{\pi^+\pi^-}_{\bar{q}}.
	\label{eq:diff-isospin}
\end{gather}
As a result, only the $u$ and $d$ quarks contribute to the azimuthal modulations in Eq.~\eqref{eq:diXsec}, providing a clean probe of their Yukawa couplings.

At the LO~\footnote{Higher-order QCD corrections are not expected to qualitatively alter our conclusions, analogous to the cancellation effects observed in spin asymmetry ratios in polarized Drell–Yan~\cite{deFlorian:2017ogw} and SIDIS~\cite{Rein:2025pwu,Rein:2025qhe}.}, the integration of the hard coefficients $\mathcal{C}_q$ and the transverse spin components can be factorized from the FFs in Eq.~\eqref{eq:diXsec}. Consequently, we can 
separately integrate over the hard-scattering variables $(x_q, x_{\bar q}, \cos\theta, \phi )$ for  object $\mathcal{O}_{\rm hard}$ and over the fragmentation variables $(z, \bar{z}, M_h)$ for $\mathcal{O}_{FF}$, which we denote as
\begin{align}
\langle \mathcal{O} \rangle &\equiv \int dx_q \, dx_{\bar q} \, d\cos\theta \, d\phi\, \mathcal{O}_{\rm hard}(x_q, x_{\bar q}, \cos\theta, \phi), \nn\\
\langle \mathcal{O} \rangle &\equiv \int dz \, d\bar{z} \, dM_h\, \mathcal{O}_{FF}(z, \bar{z}, M_h),
\end{align}
so that the hard-scattering and fragmentation contributions can be calculated independently and combined to yield the full observable,
\begin{equation}\label{eq:intXsec}
	\begin{aligned}
		& 2\pi \frac{d\sigma}{d\phi_R} 
		= \sum_{q=u,d} \, \left\langle \mathcal{C}_q  \right\rangle  \left\langle D^{h_1 h_2}_{q}  \right\rangle  \left\langle  D^{h^\prime}_{\bar{q}} + D^{h^\prime}_{q }  \right\rangle \\
		& - \big[ y_q  \sin\phi_R + \tilde{y}_q \cos\phi_R   \big] \left\langle  \omega_q \right\rangle \left\langle H^{h_1 h_2}_{q}  \right\rangle \left\langle  D^{h^\prime}_{\bar{q} } - D^{h^\prime}_{q }  \right\rangle \\
		& - \big[ \tilde{y}_q  \sin\phi_R - y_q  \cos\phi_R   \big] \left\langle  \tilde{\omega}_q  \right\rangle \left\langle H^{h_1 h_2}_{q}  \right\rangle \left\langle  D^{h^\prime}_{\bar{q} } + D^{h^\prime}_{q }  \right\rangle ,
	\end{aligned}
\end{equation}
where the isospin and charge conjugate symmetries of the DiFFs in Eq.~\eqref{eq:diff-isospin} have been applied. The choice of the tagged hadron $h^\prime$ is crucial for separating up- and down-quark Yukawa couplings. For identified charged hadrons with isospin symmetry, taking $h^\prime = h^+ + h^-$ ensures that $D^{h^\prime}_{\bar{q} } - D^{h^\prime}_{q } \simeq  0$, yet the $\phi_R$ distribution remains sensitive to the $CP$ structure, analogous to the two-body system~\cite{Yu:2023shd}, allowing a well-defined $CP$ transformation.
However, this cancellation reduces the overall sensitivity to the couplings. Therefore, in this Letter, we present predictions for $h^+$ and $h^-$ separately, focusing on the $CP$-conserving interactions and leaving the full phenomenological analysis for future work~\cite{wen:long}. In the numerical analysis, we consider the production channels with $h^\prime=\pi^\pm$, $K^\pm$, and $p/\bar{p}$. Sensitivity could be further improved by employing more differential observables with explicit $z$ dependence in the DiFFs, since the spin analyzing power increase at large $z$, while a detailed analysis is beyond the scope of this Letter.

To simplify the numerical analysis, we rewrite the single-differential $\phi_R$ distribution in Eq.~\eqref{eq:intXsec} as 
\begin{align}
	\frac{2\pi}{\sigma^{h^\prime}} \frac{d\sigma^{h'}}{d\phi_R} = 1 + S^{h^\prime}_x \sin\phi_R + S^{h^\prime}_y \cos\phi_R ,
	\label{eq:dis}
\end{align}
where $\sigma^{h^\prime}$ is the total cross section in the chosen kinematic region. The coefficients $(S^{h'}_x, S^{h'}_y)$ represent the linear combinations of the light-quark Yukawa couplings $(y_{u},y_{d})$ weighted by the relevant FFs. To extract these coefficients, we define the ``up-down" and ``left-right" asymmetries with respect to the incoming plane in Fig.~\ref{Fig:Geo},
\begin{align}
	A_{UD}^{h^\prime}
		&\,= \frac{\sigma^{h'}(\sin\phi_R>0) - \sigma^{h'}(\sin\phi_R<0)}{\sigma^{h'}(\sin\phi_R>0) + \sigma^{h'}(\sin\phi_R<0)}
			=  \frac{2}{\pi} \, S^{h'}_x,\nn\\
	A_{LR}^{h'}
		&\,= \frac{\sigma^{h'}(\cos\phi_R>0) - \sigma^{h'}(\cos\phi_R<0)}{\sigma^{h'}(\cos\phi_R>0) + \sigma^{h'}(\cos\phi_R<0)}
			= \frac{2}{\pi} \, S^{h'}_y ,
\label{eq:SSA}
\end{align}
where the $\sigma^{h'}(\sin\phi_R \gtrless 0)$ denotes the integrated cross section with $\sin\phi_R \gtrless 0$, etc. 
Since these asymmetries are defined as ratios, common systematic uncertainties cancel and can be neglected.
The statistical uncertainties of the asymmetries are
\begin{equation}
	\delta A_{LR, UD}^{h'}
	=\sqrt{\frac{1 - (A_{LR, UD}^{h'})^2}{N_{\rm events}}}
	\simeq \frac{1}{\sqrt{\sigma^{h'} \cdot \mathcal{L}}},
	\label{eq:stats}
\end{equation}
where $N_{\rm events}=\sigma^{h'} \mathcal{L}$   is the number of events after kinematic selections at a given collider energy and integrated luminosity $\mathcal{L}$. We take the approximation $A_{LR, UD}^{h'}\simeq 0$ in Eq.~\eqref{eq:stats}, which is consistent with the negligibly small light-quark masses.

Global extractions of DiFFs and single-hadron FFs are now available from several groups~\cite{Courtoy:2012ry, Radici:2015mwa, Sato:2016wqj, Bertone:2017tyb, Sato:2019yez, Moffat:2021dji, Cocuzza:2022jye, Pitonyak:2023gjx, Cocuzza:2023oam, Cocuzza:2023vqs, Gao:2024nkz, Gao:2024dbv,Mahaut:2025hie}.  
In our numerical analysis, we utilize the DiFFs from the JAM fits~\cite{Pitonyak:2023gjx, Cocuzza:2023oam, Cocuzza:2023vqs} and the LO single-hadron FFs $D^{h'}_q$ from the NNFF global analysis~\cite{Bertone:2017tyb}~\footnote{Different choices of DiFF (JAM~\cite{Cocuzza:2023vqs} and MAP~\cite{Courtoy:2012ry,Radici:2015mwa,Mahaut:2025hie}) and single-hadron FF (JAM~\cite{Moffat:2021dji}, NNFF~\cite{Bertone:2017tyb}, and NPC~\cite{Gao:2024dbv,Gao:2024nkz}) parametrizations are expected to yield consistent conclusions in this Letter.}. We estimate that the relative uncertainties from DiFFs and FFs in the spin asymmetries are below the $20\%$ level, with statistical errors being dominant. The associated uncertainties thus have negligible impact on our conclusions, and are expected to improve with forthcoming data from the EIC and future lepton colliders~\cite{Zhou:2024cyk}. We impose 
the following optimized kinematic selections in the analysis. The allowed region for the energy fractions $x_q$ and $x_{\bar{q}}$ is determined by momentum conservation and can be written as: $x_q \in (0,x_{\rm max})$, $x_{\bar{q}} \in (x_{\rm max}-x_q,1-(1-x_{\rm max})/(1-x_q))$, where $x_{\rm max}=1-m_Z^2/s$, together with the full $(\theta,\phi)$ angular range. For FFs, we impose the kinematic cuts $z \in (0.25, 0.9)$, $\bar{z} \in (0.1, 0.9)$, and $M_h \in (0.3, 2.0)~\mathrm{GeV}$, chosen to be consistent with existing global fits and to minimize uncertainties~\cite{Bertone:2017tyb,Pitonyak:2023gjx, Cocuzza:2023oam, Cocuzza:2023vqs}. Events are identified using the recoil-mass method by selecting the reconstructed $Z$-boson energy in the window $(109.2, 111.2)~{\rm GeV}$.  Since the $\sin\phi_R$ modulation for the $CP$-even $y_q$ coupling flips sign across the Higgs boson threshold, we split the $Z$-boson energy into two bins below and above the threshold to enhance sensitivity, while the $\cos\phi_R$ distribution remains unaffected. To reconstruct the $Z$ boson, we include both leptonic and hadronic decay modes, $Z\to \ell^+\ell^-/jj$ with $\ell=e,\mu, \tau$.

The projected constraints on the light-quark Yukawa couplings $(y_{u},y_{d})$ are estimated from the azimuthal asymmetries in Eq.~\eqref{eq:SSA} through a $\chi^2$ analysis, 
\begin{equation}
	\chi^2 = \sum_i \left(\frac{A_{i}^{h' , \rm th}-A_{i}^{h',\rm exp}}{\delta A_{i}^{h'}}\right)^2 \, ,
\end{equation}
where the uncertainties are dominated by the statistical errors from Eq.~\eqref{eq:stats} and the $i$ runs over all independent bins and observables to combine the constraints.

%------------------------------------------------------------------------
\begin{figure}
	\centering
	\includegraphics[scale=0.5]{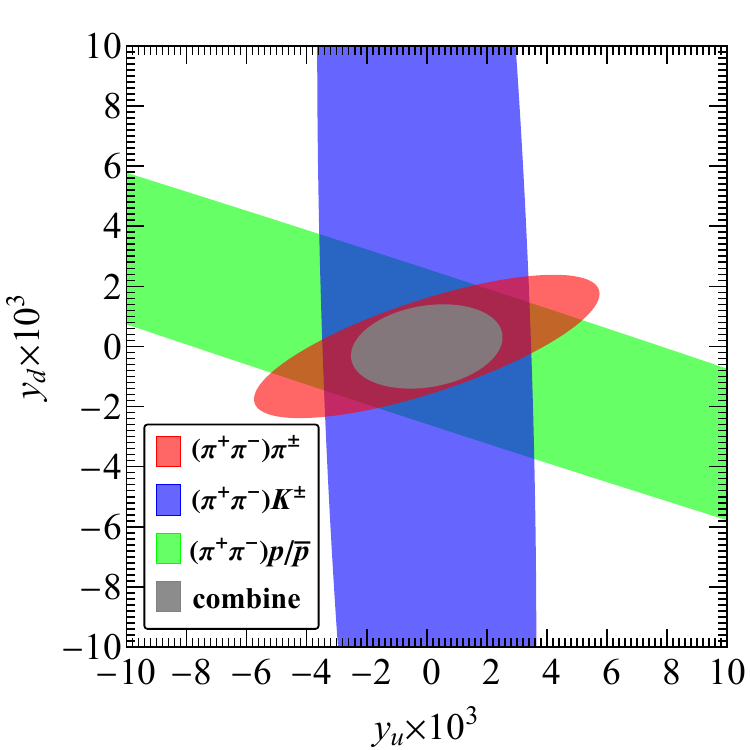}
	\caption{Expected 68\% C.L. limits on light-quark Yukawa couplings from the channels $(\pi^+\pi^-)\pi^\pm$ (red), $(\pi^+\pi^-)K^\pm$ (blue), and $(\pi^+\pi^-)p/\bar{p}$ (green), with their combination in gray, assuming $CP$ conservation and extracted from the azimuthal asymmetries $A_{LR, UD}^{h'}$.
	}
\label{Fig:Limit}
\end{figure}
%------------------------------------------------------------------------
Figure~\ref{Fig:Limit} shows the 68\% confidence level limits for the $\pi^\pm$ (red), $K^\pm$ (blue), and $p/\bar{p}$ (green) associated production channels, along with the combined result in gray. The limits indicate that the $K^\pm$ channel is primarily sensitive to the up-quark Yukawa coupling, while the $\pi^\pm$ and $p/\bar{p}$ channels yield constraint bands with opposite slopes, as determined by the combination $\langle D^{h'}_{\bar q}-D^{h'}_q\rangle$. For pions, e.g. $\pi^+(u\bar d)$, this term has opposite signs for $u$ and $d$ quarks, whereas for protons, e.g. $p(uud)$, it has the same sign for both flavors, leading to the observed opposite slopes between the $\pi^\pm$ and $p/\bar{p}$ channels.
Combining all single hadron channels is therefore essential to disentangle the flavor dependence, leading to typical upper limits of $\mathcal{O}(10^{-4}\sim 10^{-3})$, and with slightly stronger sensitivity to the down-quark Yukawa coupling. This is primarily driven by partonic hard scattering in the $\pi^\pm$ channel, while in the $p/\bar{p}$ channel it receives additional contributions from $u$-$d$ differences in FFs, further enhancing the sensitivity to $y_d$ relative to $y_u$, under the assumption of charge-conjugation and isospin symmetry~\cite{Bertone:2017tyb}.
The overall sensitivity is dominated by the $\pi^\pm$ mode because the statistical uncertainties of $A_{LR, UD}^{h'}$ for the $K^\pm$ and $p/\bar{p}$ final states remain larger. These results complement existing constraints from other approaches, which often suffer from degeneracies in flavor space and thus cannot fully resolve the underlying quark flavor structure.

%================================================================================================
\vspace{3mm}
\emph{Conclusion---}
%================================================================================================
In this Letter, we demonstrated that transverse spin dependent observables in dihadron fragmentation provide a previously unexplored avenue for probing the light-quark Yukawa couplings. The relevant azimuthal modulations arise from transverse spin effects of light quarks induced by Yukawa interactions through the interference between the Higgs induced signal and the SM background in the process $e^-e^+\to q\bar{q}Z$, followed by quark fragmentation into hadrons. These angular structures depend linearly on the Yukawa couplings $y_q$ and $\tilde{y}_q$, offering a sensitivity pattern fundamentally different from rate-based approaches, which scale as $y_q^2$. By exploiting identified single hadrons from the recoiling quark as flavor tags, we showed that the associated channels $\pi^\pm,K^\pm$, and $p/\bar{p}$ provide complementary information that cleanly separates the up- and down-quark contributions. Their combination yields stringent constraints on the flavor structure of the Higgs Yukawa sector, reaching the  $\mathcal{O}(10^{-4}\sim 10^{-3})$ level for the light-quark couplings. In contrast to the quark Boer-Mulders fracture function approach proposed in Ref.~\cite{Michel:2025afc}, our method does not rely on forward hadron identification in $pp$ collisions, thereby avoiding the associated challenges from pileup and limited detector acceptance at the LHC~\cite{LHCb:2012lfk,Soyez:2018opl,LHCb:2018roe}. These findings demonstrate that spin-dependent fragmentation offers a viable and powerful method for resolving light-quark Yukawa interactions beyond the reach of conventional probes, opening new directions for precision test of the flavor structures of the SM and for exploring potential new physics.

%================================================================================================
\vspace{3mm}
%================================================================================================
We thank C.-P. Yuan for the helpful discussion. This work is partly supported by the National Natural Science Foundation of China under Grants No.~12547174, No.~12422506, No.~12342502, and No.~12235001 and CAS under Grant No.~E429A6M1, and is partly supported by Fundamental and Interdisciplinary Disciplines Breakthrough Plan of the Ministry of Education of China-JYB2025XDXM204. The authors gratefully acknowledge the valuable discussions and insights provided by the members of the Collaboration on Precision Tests and New Physics (CPTNP).

%================================================================================================
\bibliographystyle{apsrev}
\bibliography{reference}

\begin{thebibliography}{77}
\expandafter\ifx\csname natexlab\endcsname\relax\def\natexlab#1{#1}\fi
\expandafter\ifx\csname bibnamefont\endcsname\relax
  \def\bibnamefont#1{#1}\fi
\expandafter\ifx\csname bibfnamefont\endcsname\relax
  \def\bibfnamefont#1{#1}\fi
\expandafter\ifx\csname citenamefont\endcsname\relax
  \def\citenamefont#1{#1}\fi
\expandafter\ifx\csname url\endcsname\relax
  \def\url#1{\texttt{#1}}\fi
\expandafter\ifx\csname urlprefix\endcsname\relax\def\urlprefix{URL }\fi
\providecommand{\bibinfo}[2]{#2}
\providecommand{\eprint}[2][]{\url{#2}}

\bibitem[{\citenamefont{Cao et~al.}(2017)\citenamefont{Cao, Chen, and
  Liu}}]{Cao:2016wib}
\bibinfo{author}{\bibfnamefont{Q.-H.} \bibnamefont{Cao}},
  \bibinfo{author}{\bibfnamefont{S.-L.} \bibnamefont{Chen}}, \bibnamefont{and}
  \bibinfo{author}{\bibfnamefont{Y.}~\bibnamefont{Liu}},
  \bibinfo{journal}{Phys. Rev. D} \textbf{\bibinfo{volume}{95}},
  \bibinfo{pages}{053004} (\bibinfo{year}{2017}), \eprint{1602.01934}.

\bibitem[{\citenamefont{Cao et~al.}(2019)\citenamefont{Cao, Chen, Liu, Zhang,
  and Zhang}}]{Cao:2019ygh}
\bibinfo{author}{\bibfnamefont{Q.-H.} \bibnamefont{Cao}},
  \bibinfo{author}{\bibfnamefont{S.-L.} \bibnamefont{Chen}},
  \bibinfo{author}{\bibfnamefont{Y.}~\bibnamefont{Liu}},
  \bibinfo{author}{\bibfnamefont{R.}~\bibnamefont{Zhang}}, \bibnamefont{and}
  \bibinfo{author}{\bibfnamefont{Y.}~\bibnamefont{Zhang}},
  \bibinfo{journal}{Phys. Rev. D} \textbf{\bibinfo{volume}{99}},
  \bibinfo{pages}{113003} (\bibinfo{year}{2019}), \eprint{1901.04567}.

\bibitem[{\citenamefont{Li et~al.}(2020)\citenamefont{Li, Xu, Yan, and
  Yuan}}]{Li:2019uyy}
\bibinfo{author}{\bibfnamefont{G.}~\bibnamefont{Li}},
  \bibinfo{author}{\bibfnamefont{L.-X.} \bibnamefont{Xu}},
  \bibinfo{author}{\bibfnamefont{B.}~\bibnamefont{Yan}}, \bibnamefont{and}
  \bibinfo{author}{\bibfnamefont{C.~P.} \bibnamefont{Yuan}},
  \bibinfo{journal}{Phys. Lett. B} \textbf{\bibinfo{volume}{800}},
  \bibinfo{pages}{135070} (\bibinfo{year}{2020}), \eprint{1904.12006}.

\bibitem[{\citenamefont{Cao et~al.}(2021)\citenamefont{Cao, Xie, Zhang, and
  Zhang}}]{Cao:2020hhb}
\bibinfo{author}{\bibfnamefont{Q.-H.} \bibnamefont{Cao}},
  \bibinfo{author}{\bibfnamefont{K.-P.} \bibnamefont{Xie}},
  \bibinfo{author}{\bibfnamefont{H.}~\bibnamefont{Zhang}}, \bibnamefont{and}
  \bibinfo{author}{\bibfnamefont{R.}~\bibnamefont{Zhang}},
  \bibinfo{journal}{Chin. Phys. C} \textbf{\bibinfo{volume}{45}},
  \bibinfo{pages}{023117} (\bibinfo{year}{2021}), \eprint{2008.13442}.

\bibitem[{\citenamefont{Bi et~al.}(2021)\citenamefont{Bi, Chai, Gao, Liu, and
  Zhang}}]{Bi:2020frc}
\bibinfo{author}{\bibfnamefont{Q.}~\bibnamefont{Bi}},
  \bibinfo{author}{\bibfnamefont{K.}~\bibnamefont{Chai}},
  \bibinfo{author}{\bibfnamefont{J.}~\bibnamefont{Gao}},
  \bibinfo{author}{\bibfnamefont{Y.}~\bibnamefont{Liu}}, \bibnamefont{and}
  \bibinfo{author}{\bibfnamefont{H.}~\bibnamefont{Zhang}},
  \bibinfo{journal}{Chin. Phys. C} \textbf{\bibinfo{volume}{45}},
  \bibinfo{pages}{023105} (\bibinfo{year}{2021}), \eprint{2009.02000}.

\bibitem[{\citenamefont{Aad et~al.}(2022)}]{ATLAS:2022vkf}
\bibinfo{author}{\bibfnamefont{G.}~\bibnamefont{Aad}} \bibnamefont{et~al.}
  (\bibinfo{collaboration}{ATLAS}), \bibinfo{journal}{Nature}
  \textbf{\bibinfo{volume}{607}}, \bibinfo{pages}{52} (\bibinfo{year}{2022}),
  \bibinfo{note}{[Erratum: Nature 612, E24 (2022)]}, \eprint{2207.00092}.

\bibitem[{\citenamefont{Tumasyan et~al.}(2022)}]{CMS:2022dwd}
\bibinfo{author}{\bibfnamefont{A.}~\bibnamefont{Tumasyan}} \bibnamefont{et~al.}
  (\bibinfo{collaboration}{CMS}), \bibinfo{journal}{Nature}
  \textbf{\bibinfo{volume}{607}}, \bibinfo{pages}{60} (\bibinfo{year}{2022}),
  \bibinfo{note}{[Erratum: Nature 623, (2023)]}, \eprint{2207.00043}.

\bibitem[{\citenamefont{Tumasyan et~al.}(2023)}]{CMS:2022kdi}
\bibinfo{author}{\bibfnamefont{A.}~\bibnamefont{Tumasyan}} \bibnamefont{et~al.}
  (\bibinfo{collaboration}{CMS}), \bibinfo{journal}{Eur. Phys. J. C}
  \textbf{\bibinfo{volume}{83}}, \bibinfo{pages}{562} (\bibinfo{year}{2023}),
  \eprint{2204.12957}.

\bibitem[{\citenamefont{Aad et~al.}(2025{\natexlab{a}})}]{ATLAS:2025eua}
\bibinfo{author}{\bibfnamefont{G.}~\bibnamefont{Aad}} \bibnamefont{et~al.}
  (\bibinfo{collaboration}{ATLAS}) (\bibinfo{year}{2025}{\natexlab{a}}),
  \eprint{2510.23755}.

\bibitem[{\citenamefont{Aad et~al.}(2025{\natexlab{b}})}]{ATLAS:2024wfv}
\bibinfo{author}{\bibfnamefont{G.}~\bibnamefont{Aad}} \bibnamefont{et~al.}
  (\bibinfo{collaboration}{ATLAS}), \bibinfo{journal}{JHEP}
  \textbf{\bibinfo{volume}{03}}, \bibinfo{pages}{010}
  (\bibinfo{year}{2025}{\natexlab{b}}), \eprint{2407.16320}.

\bibitem[{\citenamefont{Aad et~al.}(2025{\natexlab{c}})}]{ATLAS:2024yzu}
\bibinfo{author}{\bibfnamefont{G.}~\bibnamefont{Aad}} \bibnamefont{et~al.}
  (\bibinfo{collaboration}{ATLAS}), \bibinfo{journal}{JHEP}
  \textbf{\bibinfo{volume}{04}}, \bibinfo{pages}{075}
  (\bibinfo{year}{2025}{\natexlab{c}}), \eprint{2410.19611}.

\bibitem[{\citenamefont{Bodwin et~al.}(2013)\citenamefont{Bodwin, Petriello,
  Stoynev, and Velasco}}]{Bodwin:2013gca}
\bibinfo{author}{\bibfnamefont{G.~T.} \bibnamefont{Bodwin}},
  \bibinfo{author}{\bibfnamefont{F.}~\bibnamefont{Petriello}},
  \bibinfo{author}{\bibfnamefont{S.}~\bibnamefont{Stoynev}}, \bibnamefont{and}
  \bibinfo{author}{\bibfnamefont{M.}~\bibnamefont{Velasco}},
  \bibinfo{journal}{Phys. Rev. D} \textbf{\bibinfo{volume}{88}},
  \bibinfo{pages}{053003} (\bibinfo{year}{2013}), \eprint{1306.5770}.

\bibitem[{\citenamefont{Kagan et~al.}(2015)\citenamefont{Kagan, Perez,
  Petriello, Soreq, Stoynev, and Zupan}}]{Kagan:2014ila}
\bibinfo{author}{\bibfnamefont{A.~L.} \bibnamefont{Kagan}},
  \bibinfo{author}{\bibfnamefont{G.}~\bibnamefont{Perez}},
  \bibinfo{author}{\bibfnamefont{F.}~\bibnamefont{Petriello}},
  \bibinfo{author}{\bibfnamefont{Y.}~\bibnamefont{Soreq}},
  \bibinfo{author}{\bibfnamefont{S.}~\bibnamefont{Stoynev}}, \bibnamefont{and}
  \bibinfo{author}{\bibfnamefont{J.}~\bibnamefont{Zupan}},
  \bibinfo{journal}{Phys. Rev. Lett.} \textbf{\bibinfo{volume}{114}},
  \bibinfo{pages}{101802} (\bibinfo{year}{2015}), \eprint{1406.1722}.

\bibitem[{\citenamefont{K{\"o}nig and Neubert}(2015)}]{Konig:2015qat}
\bibinfo{author}{\bibfnamefont{M.}~\bibnamefont{K{\"o}nig}} \bibnamefont{and}
  \bibinfo{author}{\bibfnamefont{M.}~\bibnamefont{Neubert}},
  \bibinfo{journal}{JHEP} \textbf{\bibinfo{volume}{08}}, \bibinfo{pages}{012}
  (\bibinfo{year}{2015}), \eprint{1505.03870}.

\bibitem[{\citenamefont{Brivio et~al.}(2015)\citenamefont{Brivio, Goertz, and
  Isidori}}]{Brivio:2015fxa}
\bibinfo{author}{\bibfnamefont{I.}~\bibnamefont{Brivio}},
  \bibinfo{author}{\bibfnamefont{F.}~\bibnamefont{Goertz}}, \bibnamefont{and}
  \bibinfo{author}{\bibfnamefont{G.}~\bibnamefont{Isidori}},
  \bibinfo{journal}{Phys. Rev. Lett.} \textbf{\bibinfo{volume}{115}},
  \bibinfo{pages}{211801} (\bibinfo{year}{2015}), \eprint{1507.02916}.

\bibitem[{\citenamefont{Aad et~al.}(2025{\natexlab{d}})}]{ATLAS:2024ext}
\bibinfo{author}{\bibfnamefont{G.}~\bibnamefont{Aad}} \bibnamefont{et~al.}
  (\bibinfo{collaboration}{ATLAS}), \bibinfo{journal}{JHEP}
  \textbf{\bibinfo{volume}{02}}, \bibinfo{pages}{045}
  (\bibinfo{year}{2025}{\natexlab{d}}), \eprint{2407.15550}.

\bibitem[{\citenamefont{Hayrapetyan et~al.}(2025)}]{CMS:2025rxu}
\bibinfo{author}{\bibfnamefont{A.}~\bibnamefont{Hayrapetyan}}
  \bibnamefont{et~al.} (\bibinfo{collaboration}{CMS}) (\bibinfo{year}{2025}),
  \eprint{2508.14988}.

\bibitem[{\citenamefont{Aad et~al.}(2025{\natexlab{e}})}]{ATLAS:2025coj}
\bibinfo{author}{\bibfnamefont{G.}~\bibnamefont{Aad}} \bibnamefont{et~al.}
  (\bibinfo{collaboration}{ATLAS}), \bibinfo{journal}{Phys. Rev. Lett.}
  \textbf{\bibinfo{volume}{135}}, \bibinfo{pages}{231802}
  (\bibinfo{year}{2025}{\natexlab{e}}), \eprint{2507.03595}.

\bibitem[{\citenamefont{Hayrapetyan et~al.}(2024)}]{CMS:2023rcv}
\bibinfo{author}{\bibfnamefont{A.}~\bibnamefont{Hayrapetyan}}
  \bibnamefont{et~al.} (\bibinfo{collaboration}{CMS}), \bibinfo{journal}{Phys.
  Rev. Lett.} \textbf{\bibinfo{volume}{132}}, \bibinfo{pages}{121901}
  (\bibinfo{year}{2024}), \eprint{2310.05164}.

\bibitem[{\citenamefont{Chekhovsky et~al.}(2025)}]{CMS:2025xkn}
\bibinfo{author}{\bibfnamefont{V.}~\bibnamefont{Chekhovsky}}
  \bibnamefont{et~al.} (\bibinfo{collaboration}{CMS}), \bibinfo{journal}{Phys.
  Rev. D} \textbf{\bibinfo{volume}{112}}, \bibinfo{pages}{112001}
  (\bibinfo{year}{2025}), \eprint{2502.05665}.

\bibitem[{\citenamefont{Bar-Shalom and Soni}(2018)}]{Bar-Shalom:2018rjs}
\bibinfo{author}{\bibfnamefont{S.}~\bibnamefont{Bar-Shalom}} \bibnamefont{and}
  \bibinfo{author}{\bibfnamefont{A.}~\bibnamefont{Soni}},
  \bibinfo{journal}{Phys. Rev. D} \textbf{\bibinfo{volume}{98}},
  \bibinfo{pages}{055001} (\bibinfo{year}{2018}), \eprint{1804.02400}.

\bibitem[{\citenamefont{Erdelyi et~al.}(2025)\citenamefont{Erdelyi, Gr{\"o}ber,
  and Selimovic}}]{Erdelyi:2024sls}
\bibinfo{author}{\bibfnamefont{B.~A.} \bibnamefont{Erdelyi}},
  \bibinfo{author}{\bibfnamefont{R.}~\bibnamefont{Gr{\"o}ber}},
  \bibnamefont{and}
  \bibinfo{author}{\bibfnamefont{N.}~\bibnamefont{Selimovic}},
  \bibinfo{journal}{JHEP} \textbf{\bibinfo{volume}{05}}, \bibinfo{pages}{189}
  (\bibinfo{year}{2025}), \eprint{2410.08272}.

\bibitem[{\citenamefont{Zhou}(2016)}]{Zhou:2015wra}
\bibinfo{author}{\bibfnamefont{Y.}~\bibnamefont{Zhou}}, \bibinfo{journal}{Phys.
  Rev. D} \textbf{\bibinfo{volume}{93}}, \bibinfo{pages}{013019}
  (\bibinfo{year}{2016}), \eprint{1505.06369}.

\bibitem[{\citenamefont{Bishara et~al.}(2017)\citenamefont{Bishara, Haisch,
  Monni, and Re}}]{Bishara:2016jga}
\bibinfo{author}{\bibfnamefont{F.}~\bibnamefont{Bishara}},
  \bibinfo{author}{\bibfnamefont{U.}~\bibnamefont{Haisch}},
  \bibinfo{author}{\bibfnamefont{P.~F.} \bibnamefont{Monni}}, \bibnamefont{and}
  \bibinfo{author}{\bibfnamefont{E.}~\bibnamefont{Re}}, \bibinfo{journal}{Phys.
  Rev. Lett.} \textbf{\bibinfo{volume}{118}}, \bibinfo{pages}{121801}
  (\bibinfo{year}{2017}), \eprint{1606.09253}.

\bibitem[{\citenamefont{Soreq et~al.}(2016)\citenamefont{Soreq, Zhu, and
  Zupan}}]{Soreq:2016rae}
\bibinfo{author}{\bibfnamefont{Y.}~\bibnamefont{Soreq}},
  \bibinfo{author}{\bibfnamefont{H.~X.} \bibnamefont{Zhu}}, \bibnamefont{and}
  \bibinfo{author}{\bibfnamefont{J.}~\bibnamefont{Zupan}},
  \bibinfo{journal}{JHEP} \textbf{\bibinfo{volume}{12}}, \bibinfo{pages}{045}
  (\bibinfo{year}{2016}), \eprint{1606.09621}.

\bibitem[{\citenamefont{Gao}(2018)}]{Gao:2016jcm}
\bibinfo{author}{\bibfnamefont{J.}~\bibnamefont{Gao}}, \bibinfo{journal}{JHEP}
  \textbf{\bibinfo{volume}{01}}, \bibinfo{pages}{038} (\bibinfo{year}{2018}),
  \eprint{1608.01746}.

\bibitem[{\citenamefont{Yu}(2017)}]{Yu:2016rvv}
\bibinfo{author}{\bibfnamefont{F.}~\bibnamefont{Yu}}, \bibinfo{journal}{JHEP}
  \textbf{\bibinfo{volume}{02}}, \bibinfo{pages}{083} (\bibinfo{year}{2017}),
  \eprint{1609.06592}.

\bibitem[{\citenamefont{Falkowski et~al.}(2021)\citenamefont{Falkowski,
  Ganguly, Gras, No, Tobioka, Vignaroli, and You}}]{Falkowski:2020znk}
\bibinfo{author}{\bibfnamefont{A.}~\bibnamefont{Falkowski}},
  \bibinfo{author}{\bibfnamefont{S.}~\bibnamefont{Ganguly}},
  \bibinfo{author}{\bibfnamefont{P.}~\bibnamefont{Gras}},
  \bibinfo{author}{\bibfnamefont{J.~M.} \bibnamefont{No}},
  \bibinfo{author}{\bibfnamefont{K.}~\bibnamefont{Tobioka}},
  \bibinfo{author}{\bibfnamefont{N.}~\bibnamefont{Vignaroli}},
  \bibnamefont{and} \bibinfo{author}{\bibfnamefont{T.}~\bibnamefont{You}},
  \bibinfo{journal}{JHEP} \textbf{\bibinfo{volume}{04}}, \bibinfo{pages}{023}
  (\bibinfo{year}{2021}), \eprint{2011.09551}.

\bibitem[{\citenamefont{Vignaroli}(2022)}]{Vignaroli:2022fqh}
\bibinfo{author}{\bibfnamefont{N.}~\bibnamefont{Vignaroli}},
  \bibinfo{journal}{Symmetry} \textbf{\bibinfo{volume}{14}},
  \bibinfo{pages}{1183} (\bibinfo{year}{2022}), \eprint{2205.09449}.

\bibitem[{\citenamefont{Alasfar et~al.}(2022)\citenamefont{Alasfar, Gr{\"o}ber,
  Grojean, Paul, and Qian}}]{Alasfar:2022vqw}
\bibinfo{author}{\bibfnamefont{L.}~\bibnamefont{Alasfar}},
  \bibinfo{author}{\bibfnamefont{R.}~\bibnamefont{Gr{\"o}ber}},
  \bibinfo{author}{\bibfnamefont{C.}~\bibnamefont{Grojean}},
  \bibinfo{author}{\bibfnamefont{A.}~\bibnamefont{Paul}}, \bibnamefont{and}
  \bibinfo{author}{\bibfnamefont{Z.}~\bibnamefont{Qian}},
  \bibinfo{journal}{JHEP} \textbf{\bibinfo{volume}{11}}, \bibinfo{pages}{045}
  (\bibinfo{year}{2022}), \eprint{2207.04157}.

\bibitem[{\citenamefont{Balzani et~al.}(2023)\citenamefont{Balzani, Gr{\"o}ber,
  and Vitti}}]{Balzani:2023jas}
\bibinfo{author}{\bibfnamefont{E.}~\bibnamefont{Balzani}},
  \bibinfo{author}{\bibfnamefont{R.}~\bibnamefont{Gr{\"o}ber}},
  \bibnamefont{and} \bibinfo{author}{\bibfnamefont{M.}~\bibnamefont{Vitti}},
  \bibinfo{journal}{JHEP} \textbf{\bibinfo{volume}{10}}, \bibinfo{pages}{027}
  (\bibinfo{year}{2023}), \eprint{2304.09772}.

\bibitem[{\citenamefont{Yan and Lee}(2024)}]{Yan:2023xsd}
\bibinfo{author}{\bibfnamefont{B.}~\bibnamefont{Yan}} \bibnamefont{and}
  \bibinfo{author}{\bibfnamefont{C.}~\bibnamefont{Lee}},
  \bibinfo{journal}{JHEP} \textbf{\bibinfo{volume}{03}}, \bibinfo{pages}{123}
  (\bibinfo{year}{2024}), \eprint{2311.12556}.

\bibitem[{\citenamefont{Michel}(2025)}]{Michel:2025afc}
\bibinfo{author}{\bibfnamefont{J.~K.~L.} \bibnamefont{Michel}}
  (\bibinfo{year}{2025}), \eprint{2508.05914}.

\bibitem[{\citenamefont{de~Blas et~al.}(2020)}]{deBlas:2019rxi}
\bibinfo{author}{\bibfnamefont{J.}~\bibnamefont{de~Blas}} \bibnamefont{et~al.},
  \bibinfo{journal}{JHEP} \textbf{\bibinfo{volume}{01}}, \bibinfo{pages}{139}
  (\bibinfo{year}{2020}), \eprint{1905.03764}.

\bibitem[{\citenamefont{Collins et~al.}(1994)\citenamefont{Collins, Heppelmann,
  and Ladinsky}}]{Collins:1993kq}
\bibinfo{author}{\bibfnamefont{J.~C.} \bibnamefont{Collins}},
  \bibinfo{author}{\bibfnamefont{S.~F.} \bibnamefont{Heppelmann}},
  \bibnamefont{and} \bibinfo{author}{\bibfnamefont{G.~A.}
  \bibnamefont{Ladinsky}}, \bibinfo{journal}{Nucl. Phys. B}
  \textbf{\bibinfo{volume}{420}}, \bibinfo{pages}{565} (\bibinfo{year}{1994}),
  \eprint{hep-ph/9305309}.

\bibitem[{\citenamefont{Collins}(2011)}]{Collins:2011zzd}
\bibinfo{author}{\bibfnamefont{J.}~\bibnamefont{Collins}},
  \emph{\bibinfo{title}{{Foundations of Perturbative QCD}}},
  vol.~\bibinfo{volume}{32} (\bibinfo{publisher}{Cambridge University Press},
  \bibinfo{year}{2011}), ISBN \bibinfo{isbn}{978-1-009-40184-5,
  978-1-009-40183-8, 978-1-009-40182-1}.

\bibitem[{\citenamefont{Zhou and Metz}(2011)}]{Zhou:2011ba}
\bibinfo{author}{\bibfnamefont{J.}~\bibnamefont{Zhou}} \bibnamefont{and}
  \bibinfo{author}{\bibfnamefont{A.}~\bibnamefont{Metz}},
  \bibinfo{journal}{Phys. Rev. Lett.} \textbf{\bibinfo{volume}{106}},
  \bibinfo{pages}{172001} (\bibinfo{year}{2011}), \eprint{1101.3273}.

\bibitem[{\citenamefont{Boughezal et~al.}(2023)\citenamefont{Boughezal,
  de~Florian, Petriello, and Vogelsang}}]{Boughezal:2023ooo}
\bibinfo{author}{\bibfnamefont{R.}~\bibnamefont{Boughezal}},
  \bibinfo{author}{\bibfnamefont{D.}~\bibnamefont{de~Florian}},
  \bibinfo{author}{\bibfnamefont{F.}~\bibnamefont{Petriello}},
  \bibnamefont{and}
  \bibinfo{author}{\bibfnamefont{W.}~\bibnamefont{Vogelsang}},
  \bibinfo{journal}{Phys. Rev. D} \textbf{\bibinfo{volume}{107}},
  \bibinfo{pages}{075028} (\bibinfo{year}{2023}), \eprint{2301.02304}.

\bibitem[{\citenamefont{Wang et~al.}(2024)\citenamefont{Wang, Wen, Xing, and
  Yan}}]{Wang:2024zns}
\bibinfo{author}{\bibfnamefont{H.-L.} \bibnamefont{Wang}},
  \bibinfo{author}{\bibfnamefont{X.-K.} \bibnamefont{Wen}},
  \bibinfo{author}{\bibfnamefont{H.}~\bibnamefont{Xing}}, \bibnamefont{and}
  \bibinfo{author}{\bibfnamefont{B.}~\bibnamefont{Yan}},
  \bibinfo{journal}{Phys. Rev. D} \textbf{\bibinfo{volume}{109}},
  \bibinfo{pages}{095025} (\bibinfo{year}{2024}), \eprint{2401.08419}.

\bibitem[{\citenamefont{Wen et~al.}(2024)\citenamefont{Wen, Yan, Yu, and
  Yuan}}]{Wen:2024cfu}
\bibinfo{author}{\bibfnamefont{X.-K.} \bibnamefont{Wen}},
  \bibinfo{author}{\bibfnamefont{B.}~\bibnamefont{Yan}},
  \bibinfo{author}{\bibfnamefont{Z.}~\bibnamefont{Yu}}, \bibnamefont{and}
  \bibinfo{author}{\bibfnamefont{C.~P.} \bibnamefont{Yuan}}
  (\bibinfo{year}{2024}), \eprint{2408.07255}.

\bibitem[{\citenamefont{Wen et~al.}(2025)\citenamefont{Wen, Yan, Yu, and
  Yuan}}]{Wen:2024nff}
\bibinfo{author}{\bibfnamefont{X.-K.} \bibnamefont{Wen}},
  \bibinfo{author}{\bibfnamefont{B.}~\bibnamefont{Yan}},
  \bibinfo{author}{\bibfnamefont{Z.}~\bibnamefont{Yu}}, \bibnamefont{and}
  \bibinfo{author}{\bibfnamefont{C.~P.} \bibnamefont{Yuan}},
  \bibinfo{journal}{Phys. Rev. D} \textbf{\bibinfo{volume}{112}},
  \bibinfo{pages}{053004} (\bibinfo{year}{2025}), \eprint{2411.13845}.

\bibitem[{\citenamefont{Yang et~al.}(2025)\citenamefont{Yang, Song, and
  Wei}}]{Yang:2024kjn}
\bibinfo{author}{\bibfnamefont{L.}~\bibnamefont{Yang}},
  \bibinfo{author}{\bibfnamefont{Y.-K.} \bibnamefont{Song}}, \bibnamefont{and}
  \bibinfo{author}{\bibfnamefont{S.-Y.} \bibnamefont{Wei}},
  \bibinfo{journal}{Phys. Rev. D} \textbf{\bibinfo{volume}{111}},
  \bibinfo{pages}{054035} (\bibinfo{year}{2025}), \eprint{2410.20917}.

\bibitem[{\citenamefont{Cheng and Yan}(2025)}]{Cheng:2025cuv}
\bibinfo{author}{\bibfnamefont{K.}~\bibnamefont{Cheng}} \bibnamefont{and}
  \bibinfo{author}{\bibfnamefont{B.}~\bibnamefont{Yan}},
  \bibinfo{journal}{Phys. Rev. Lett.} \textbf{\bibinfo{volume}{135}},
  \bibinfo{pages}{011902} (\bibinfo{year}{2025}), \eprint{2501.03321}.

\bibitem[{\citenamefont{Huang et~al.}(2025)\citenamefont{Huang, Tong, and
  Wang}}]{Huang:2025ljp}
\bibinfo{author}{\bibfnamefont{Y.}~\bibnamefont{Huang}},
  \bibinfo{author}{\bibfnamefont{X.-B.} \bibnamefont{Tong}}, \bibnamefont{and}
  \bibinfo{author}{\bibfnamefont{H.-L.} \bibnamefont{Wang}}
  (\bibinfo{year}{2025}), \eprint{2508.08516}.

\bibitem[{\citenamefont{Cao et~al.}(2025)\citenamefont{Cao, Li, Wen, and
  Yan}}]{Cao:2025qua}
\bibinfo{author}{\bibfnamefont{Q.-H.} \bibnamefont{Cao}},
  \bibinfo{author}{\bibfnamefont{G.}~\bibnamefont{Li}},
  \bibinfo{author}{\bibfnamefont{X.-K.} \bibnamefont{Wen}}, \bibnamefont{and}
  \bibinfo{author}{\bibfnamefont{B.}~\bibnamefont{Yan}} (\bibinfo{year}{2025}),
  \eprint{2509.18276}.

\bibitem[{\citenamefont{Qin et~al.}(2025)\citenamefont{Qin, Song, and
  Wei}}]{Qin:2025cvp}
\bibinfo{author}{\bibfnamefont{X.-y.} \bibnamefont{Qin}},
  \bibinfo{author}{\bibfnamefont{Y.-K.} \bibnamefont{Song}}, \bibnamefont{and}
  \bibinfo{author}{\bibfnamefont{S.-y.} \bibnamefont{Wei}}
  (\bibinfo{year}{2025}), \eprint{2504.00739}.

\bibitem[{\citenamefont{He et~al.}(2026)\citenamefont{He, Li, Tian, Wen, and
  Yan}}]{He:2026pxa}
\bibinfo{author}{\bibfnamefont{Z.-G.} \bibnamefont{He}},
  \bibinfo{author}{\bibfnamefont{G.}~\bibnamefont{Li}},
  \bibinfo{author}{\bibfnamefont{Y.-J.} \bibnamefont{Tian}},
  \bibinfo{author}{\bibfnamefont{X.-K.} \bibnamefont{Wen}}, \bibnamefont{and}
  \bibinfo{author}{\bibfnamefont{B.}~\bibnamefont{Yan}} (\bibinfo{year}{2026}),
  \eprint{2603.18874}.

\bibitem[{\citenamefont{Wen et~al.}(2026)\citenamefont{Wen, Yan, and
  Zhang}}]{wen:long}
\bibinfo{author}{\bibfnamefont{X.-K.} \bibnamefont{Wen}},
  \bibinfo{author}{\bibfnamefont{B.}~\bibnamefont{Yan}}, \bibnamefont{and}
  \bibinfo{author}{\bibfnamefont{S.-T.} \bibnamefont{Zhang}}
  (\bibinfo{year}{2026}), \eprint{in preparation}.

\bibitem[{\citenamefont{Yu}(2023)}]{Yu:2023shd}
\bibinfo{author}{\bibfnamefont{Z.}~\bibnamefont{Yu}}, Ph.D. thesis
  (\bibinfo{year}{2023}), \eprint{2308.13080}.

\bibitem[{\citenamefont{Berman et~al.}(1971)\citenamefont{Berman, Bjorken, and
  Kogut}}]{Berman:1971xz}
\bibinfo{author}{\bibfnamefont{S.~M.} \bibnamefont{Berman}},
  \bibinfo{author}{\bibfnamefont{J.~D.} \bibnamefont{Bjorken}},
  \bibnamefont{and} \bibinfo{author}{\bibfnamefont{J.~B.} \bibnamefont{Kogut}},
  \bibinfo{journal}{Phys. Rev. D} \textbf{\bibinfo{volume}{4}},
  \bibinfo{pages}{3388} (\bibinfo{year}{1971}).

\bibitem[{\citenamefont{Field and Feynman}(1978)}]{Field:1977fa}
\bibinfo{author}{\bibfnamefont{R.~D.} \bibnamefont{Field}} \bibnamefont{and}
  \bibinfo{author}{\bibfnamefont{R.~P.} \bibnamefont{Feynman}},
  \bibinfo{journal}{Nucl. Phys. B} \textbf{\bibinfo{volume}{136}},
  \bibinfo{pages}{1} (\bibinfo{year}{1978}).

\bibitem[{\citenamefont{Feynman et~al.}(1978)\citenamefont{Feynman, Field, and
  Fox}}]{Feynman:1978dt}
\bibinfo{author}{\bibfnamefont{R.~P.} \bibnamefont{Feynman}},
  \bibinfo{author}{\bibfnamefont{R.~D.} \bibnamefont{Field}}, \bibnamefont{and}
  \bibinfo{author}{\bibfnamefont{G.~C.} \bibnamefont{Fox}},
  \bibinfo{journal}{Phys. Rev. D} \textbf{\bibinfo{volume}{18}},
  \bibinfo{pages}{3320} (\bibinfo{year}{1978}).

\bibitem[{\citenamefont{Ai et~al.}(2025)}]{Ai:2025cpj}
\bibinfo{author}{\bibfnamefont{X.}~\bibnamefont{Ai}} \bibnamefont{et~al.},
  \bibinfo{journal}{Chin. Phys. C} \textbf{\bibinfo{volume}{49}},
  \bibinfo{pages}{123108} (\bibinfo{year}{2025}), \eprint{2505.24810}.

\bibitem[{\citenamefont{Dong et~al.}(2018)}]{CEPCStudyGroup:2018ghi}
\bibinfo{author}{\bibfnamefont{M.}~\bibnamefont{Dong}} \bibnamefont{et~al.}
  (\bibinfo{collaboration}{CEPC Study Group}) (\bibinfo{year}{2018}),
  \eprint{1811.10545}.

\bibitem[{\citenamefont{Abdallah et~al.}(2024)}]{CEPCStudyGroup:2023quu}
\bibinfo{author}{\bibfnamefont{W.}~\bibnamefont{Abdallah}} \bibnamefont{et~al.}
  (\bibinfo{collaboration}{CEPC Study Group}), \bibinfo{journal}{Radiat.
  Detect. Technol. Methods} \textbf{\bibinfo{volume}{8}}, \bibinfo{pages}{1}
  (\bibinfo{year}{2024}), \bibinfo{note}{[Erratum:
  Radiat.Detect.Technol.Methods 9, 184--192 (2025)]}, \eprint{2312.14363}.

\bibitem[{ILC(2013)}]{ILC:2013jhg}
 (\bibinfo{year}{2013}), \eprint{1306.6352}.

\bibitem[{\citenamefont{Abada et~al.}(2019)}]{FCC:2018evy}
\bibinfo{author}{\bibfnamefont{A.}~\bibnamefont{Abada}} \bibnamefont{et~al.}
  (\bibinfo{collaboration}{FCC}), \bibinfo{journal}{Eur. Phys. J. ST}
  \textbf{\bibinfo{volume}{228}}, \bibinfo{pages}{261} (\bibinfo{year}{2019}).

\bibitem[{\citenamefont{Pitonyak et~al.}(2024)\citenamefont{Pitonyak, Cocuzza,
  Metz, Prokudin, and Sato}}]{Pitonyak:2023gjx}
\bibinfo{author}{\bibfnamefont{D.}~\bibnamefont{Pitonyak}},
  \bibinfo{author}{\bibfnamefont{C.}~\bibnamefont{Cocuzza}},
  \bibinfo{author}{\bibfnamefont{A.}~\bibnamefont{Metz}},
  \bibinfo{author}{\bibfnamefont{A.}~\bibnamefont{Prokudin}}, \bibnamefont{and}
  \bibinfo{author}{\bibfnamefont{N.}~\bibnamefont{Sato}},
  \bibinfo{journal}{Phys. Rev. Lett.} \textbf{\bibinfo{volume}{132}},
  \bibinfo{pages}{011902} (\bibinfo{year}{2024}), \eprint{2305.11995}.

\bibitem[{\citenamefont{Cocuzza
  et~al.}(2024{\natexlab{a}})\citenamefont{Cocuzza, Metz, Pitonyak, Prokudin,
  Sato, and Seidl}}]{Cocuzza:2023oam}
\bibinfo{author}{\bibfnamefont{C.}~\bibnamefont{Cocuzza}},
  \bibinfo{author}{\bibfnamefont{A.}~\bibnamefont{Metz}},
  \bibinfo{author}{\bibfnamefont{D.}~\bibnamefont{Pitonyak}},
  \bibinfo{author}{\bibfnamefont{A.}~\bibnamefont{Prokudin}},
  \bibinfo{author}{\bibfnamefont{N.}~\bibnamefont{Sato}}, \bibnamefont{and}
  \bibinfo{author}{\bibfnamefont{R.}~\bibnamefont{Seidl}}
  (\bibinfo{collaboration}{JAM}), \bibinfo{journal}{Phys. Rev. Lett.}
  \textbf{\bibinfo{volume}{132}}, \bibinfo{pages}{091901}
  (\bibinfo{year}{2024}{\natexlab{a}}), \eprint{2306.12998}.

\bibitem[{\citenamefont{Cocuzza
  et~al.}(2024{\natexlab{b}})\citenamefont{Cocuzza, Metz, Pitonyak, Prokudin,
  Sato, and Seidl}}]{Cocuzza:2023vqs}
\bibinfo{author}{\bibfnamefont{C.}~\bibnamefont{Cocuzza}},
  \bibinfo{author}{\bibfnamefont{A.}~\bibnamefont{Metz}},
  \bibinfo{author}{\bibfnamefont{D.}~\bibnamefont{Pitonyak}},
  \bibinfo{author}{\bibfnamefont{A.}~\bibnamefont{Prokudin}},
  \bibinfo{author}{\bibfnamefont{N.}~\bibnamefont{Sato}}, \bibnamefont{and}
  \bibinfo{author}{\bibfnamefont{R.}~\bibnamefont{Seidl}}
  (\bibinfo{collaboration}{Jefferson Lab Angular Momentum (JAM)}),
  \bibinfo{journal}{Phys. Rev. D} \textbf{\bibinfo{volume}{109}},
  \bibinfo{pages}{034024} (\bibinfo{year}{2024}{\natexlab{b}}),
  \eprint{2308.14857}.

\bibitem[{\citenamefont{de~Florian}(2017)}]{deFlorian:2017ogw}
\bibinfo{author}{\bibfnamefont{D.}~\bibnamefont{de~Florian}},
  \bibinfo{journal}{Phys. Rev. D} \textbf{\bibinfo{volume}{96}},
  \bibinfo{pages}{094006} (\bibinfo{year}{2017}), \eprint{1711.01235}.

\bibitem[{\citenamefont{Rein et~al.}(2025{\natexlab{a}})\citenamefont{Rein,
  Schlegel, Tollk{\"u}hn, and Vogelsang}}]{Rein:2025pwu}
\bibinfo{author}{\bibfnamefont{D.}~\bibnamefont{Rein}},
  \bibinfo{author}{\bibfnamefont{M.}~\bibnamefont{Schlegel}},
  \bibinfo{author}{\bibfnamefont{P.}~\bibnamefont{Tollk{\"u}hn}},
  \bibnamefont{and}
  \bibinfo{author}{\bibfnamefont{W.}~\bibnamefont{Vogelsang}},
  \bibinfo{journal}{Phys. Rev. Lett.} \textbf{\bibinfo{volume}{135}},
  \bibinfo{pages}{251901} (\bibinfo{year}{2025}{\natexlab{a}}),
  \eprint{2503.16097}.

\bibitem[{\citenamefont{Rein et~al.}(2025{\natexlab{b}})\citenamefont{Rein,
  Schlegel, Tollk{\"u}hn, and Vogelsang}}]{Rein:2025qhe}
\bibinfo{author}{\bibfnamefont{D.}~\bibnamefont{Rein}},
  \bibinfo{author}{\bibfnamefont{M.}~\bibnamefont{Schlegel}},
  \bibinfo{author}{\bibfnamefont{P.}~\bibnamefont{Tollk{\"u}hn}},
  \bibnamefont{and}
  \bibinfo{author}{\bibfnamefont{W.}~\bibnamefont{Vogelsang}},
  \bibinfo{journal}{Phys. Rev. D} \textbf{\bibinfo{volume}{112}},
  \bibinfo{pages}{114024} (\bibinfo{year}{2025}{\natexlab{b}}),
  \eprint{2503.16119}.

\bibitem[{\citenamefont{Courtoy et~al.}(2012)\citenamefont{Courtoy, Bacchetta,
  Radici, and Bianconi}}]{Courtoy:2012ry}
\bibinfo{author}{\bibfnamefont{A.}~\bibnamefont{Courtoy}},
  \bibinfo{author}{\bibfnamefont{A.}~\bibnamefont{Bacchetta}},
  \bibinfo{author}{\bibfnamefont{M.}~\bibnamefont{Radici}}, \bibnamefont{and}
  \bibinfo{author}{\bibfnamefont{A.}~\bibnamefont{Bianconi}},
  \bibinfo{journal}{Phys. Rev. D} \textbf{\bibinfo{volume}{85}},
  \bibinfo{pages}{114023} (\bibinfo{year}{2012}), \eprint{1202.0323}.

\bibitem[{\citenamefont{Radici et~al.}(2015)\citenamefont{Radici, Courtoy,
  Bacchetta, and Guagnelli}}]{Radici:2015mwa}
\bibinfo{author}{\bibfnamefont{M.}~\bibnamefont{Radici}},
  \bibinfo{author}{\bibfnamefont{A.}~\bibnamefont{Courtoy}},
  \bibinfo{author}{\bibfnamefont{A.}~\bibnamefont{Bacchetta}},
  \bibnamefont{and}
  \bibinfo{author}{\bibfnamefont{M.}~\bibnamefont{Guagnelli}},
  \bibinfo{journal}{JHEP} \textbf{\bibinfo{volume}{05}}, \bibinfo{pages}{123}
  (\bibinfo{year}{2015}), \eprint{1503.03495}.

\bibitem[{\citenamefont{Sato et~al.}(2016)\citenamefont{Sato, Ethier,
  Melnitchouk, Hirai, Kumano, and Accardi}}]{Sato:2016wqj}
\bibinfo{author}{\bibfnamefont{N.}~\bibnamefont{Sato}},
  \bibinfo{author}{\bibfnamefont{J.~J.} \bibnamefont{Ethier}},
  \bibinfo{author}{\bibfnamefont{W.}~\bibnamefont{Melnitchouk}},
  \bibinfo{author}{\bibfnamefont{M.}~\bibnamefont{Hirai}},
  \bibinfo{author}{\bibfnamefont{S.}~\bibnamefont{Kumano}}, \bibnamefont{and}
  \bibinfo{author}{\bibfnamefont{A.}~\bibnamefont{Accardi}},
  \bibinfo{journal}{Phys. Rev. D} \textbf{\bibinfo{volume}{94}},
  \bibinfo{pages}{114004} (\bibinfo{year}{2016}), \eprint{1609.00899}.

\bibitem[{\citenamefont{Bertone et~al.}(2017)\citenamefont{Bertone, Carrazza,
  Hartland, Nocera, and Rojo}}]{Bertone:2017tyb}
\bibinfo{author}{\bibfnamefont{V.}~\bibnamefont{Bertone}},
  \bibinfo{author}{\bibfnamefont{S.}~\bibnamefont{Carrazza}},
  \bibinfo{author}{\bibfnamefont{N.~P.} \bibnamefont{Hartland}},
  \bibinfo{author}{\bibfnamefont{E.~R.} \bibnamefont{Nocera}},
  \bibnamefont{and} \bibinfo{author}{\bibfnamefont{J.}~\bibnamefont{Rojo}}
  (\bibinfo{collaboration}{NNPDF}), \bibinfo{journal}{Eur. Phys. J. C}
  \textbf{\bibinfo{volume}{77}}, \bibinfo{pages}{516} (\bibinfo{year}{2017}),
  \eprint{1706.07049}.

\bibitem[{\citenamefont{Sato et~al.}(2020)\citenamefont{Sato, Andres, Ethier,
  and Melnitchouk}}]{Sato:2019yez}
\bibinfo{author}{\bibfnamefont{N.}~\bibnamefont{Sato}},
  \bibinfo{author}{\bibfnamefont{C.}~\bibnamefont{Andres}},
  \bibinfo{author}{\bibfnamefont{J.~J.} \bibnamefont{Ethier}},
  \bibnamefont{and}
  \bibinfo{author}{\bibfnamefont{W.}~\bibnamefont{Melnitchouk}}
  (\bibinfo{collaboration}{JAM}), \bibinfo{journal}{Phys. Rev. D}
  \textbf{\bibinfo{volume}{101}}, \bibinfo{pages}{074020}
  (\bibinfo{year}{2020}), \eprint{1905.03788}.

\bibitem[{\citenamefont{Moffat et~al.}(2021)\citenamefont{Moffat, Melnitchouk,
  Rogers, and Sato}}]{Moffat:2021dji}
\bibinfo{author}{\bibfnamefont{E.}~\bibnamefont{Moffat}},
  \bibinfo{author}{\bibfnamefont{W.}~\bibnamefont{Melnitchouk}},
  \bibinfo{author}{\bibfnamefont{T.~C.} \bibnamefont{Rogers}},
  \bibnamefont{and} \bibinfo{author}{\bibfnamefont{N.}~\bibnamefont{Sato}}
  (\bibinfo{collaboration}{Jefferson Lab Angular Momentum (JAM)}),
  \bibinfo{journal}{Phys. Rev. D} \textbf{\bibinfo{volume}{104}},
  \bibinfo{pages}{016015} (\bibinfo{year}{2021}), \eprint{2101.04664}.

\bibitem[{\citenamefont{Cocuzza et~al.}(2022)\citenamefont{Cocuzza,
  Melnitchouk, Metz, and Sato}}]{Cocuzza:2022jye}
\bibinfo{author}{\bibfnamefont{C.}~\bibnamefont{Cocuzza}},
  \bibinfo{author}{\bibfnamefont{W.}~\bibnamefont{Melnitchouk}},
  \bibinfo{author}{\bibfnamefont{A.}~\bibnamefont{Metz}}, \bibnamefont{and}
  \bibinfo{author}{\bibfnamefont{N.}~\bibnamefont{Sato}}
  (\bibinfo{collaboration}{Jefferson Lab Angular Momentum (JAM)}),
  \bibinfo{journal}{Phys. Rev. D} \textbf{\bibinfo{volume}{106}},
  \bibinfo{pages}{L031502} (\bibinfo{year}{2022}), \eprint{2202.03372}.

\bibitem[{\citenamefont{Gao et~al.}(2024{\natexlab{a}})\citenamefont{Gao, Liu,
  Shen, Xing, and Zhao}}]{Gao:2024nkz}
\bibinfo{author}{\bibfnamefont{J.}~\bibnamefont{Gao}},
  \bibinfo{author}{\bibfnamefont{C.}~\bibnamefont{Liu}},
  \bibinfo{author}{\bibfnamefont{X.}~\bibnamefont{Shen}},
  \bibinfo{author}{\bibfnamefont{H.}~\bibnamefont{Xing}}, \bibnamefont{and}
  \bibinfo{author}{\bibfnamefont{Y.}~\bibnamefont{Zhao}},
  \bibinfo{journal}{Phys. Rev. Lett.} \textbf{\bibinfo{volume}{132}},
  \bibinfo{pages}{261903} (\bibinfo{year}{2024}{\natexlab{a}}),
  \eprint{2401.02781}.

\bibitem[{\citenamefont{Gao et~al.}(2024{\natexlab{b}})\citenamefont{Gao, Liu,
  Shen, Xing, and Zhao}}]{Gao:2024dbv}
\bibinfo{author}{\bibfnamefont{J.}~\bibnamefont{Gao}},
  \bibinfo{author}{\bibfnamefont{C.}~\bibnamefont{Liu}},
  \bibinfo{author}{\bibfnamefont{X.}~\bibnamefont{Shen}},
  \bibinfo{author}{\bibfnamefont{H.}~\bibnamefont{Xing}}, \bibnamefont{and}
  \bibinfo{author}{\bibfnamefont{Y.}~\bibnamefont{Zhao}},
  \bibinfo{journal}{Phys. Rev. D} \textbf{\bibinfo{volume}{110}},
  \bibinfo{pages}{114019} (\bibinfo{year}{2024}{\natexlab{b}}),
  \eprint{2407.04422}.

\bibitem[{\citenamefont{Mahaut et~al.}(2025)\citenamefont{Mahaut, Polano,
  Bacchetta, Bertone, Cerutti, Radici, and Rossi}}]{Mahaut:2025hie}
\bibinfo{author}{\bibfnamefont{V.}~\bibnamefont{Mahaut}},
  \bibinfo{author}{\bibfnamefont{L.}~\bibnamefont{Polano}},
  \bibinfo{author}{\bibfnamefont{A.}~\bibnamefont{Bacchetta}},
  \bibinfo{author}{\bibfnamefont{V.}~\bibnamefont{Bertone}},
  \bibinfo{author}{\bibfnamefont{M.}~\bibnamefont{Cerutti}},
  \bibinfo{author}{\bibfnamefont{M.}~\bibnamefont{Radici}}, \bibnamefont{and}
  \bibinfo{author}{\bibfnamefont{L.}~\bibnamefont{Rossi}}
  (\bibinfo{year}{2025}), \eprint{2509.11855}.

\bibitem[{\citenamefont{Zhou and Gao}(2025)}]{Zhou:2024cyk}
\bibinfo{author}{\bibfnamefont{B.}~\bibnamefont{Zhou}} \bibnamefont{and}
  \bibinfo{author}{\bibfnamefont{J.}~\bibnamefont{Gao}},
  \bibinfo{journal}{JHEP} \textbf{\bibinfo{volume}{02}}, \bibinfo{pages}{003}
  (\bibinfo{year}{2025}), \eprint{2407.10059}.

\bibitem[{\citenamefont{Aaij et~al.}(2012)}]{LHCb:2012lfk}
\bibinfo{author}{\bibfnamefont{R.}~\bibnamefont{Aaij}} \bibnamefont{et~al.}
  (\bibinfo{collaboration}{LHCb}), \bibinfo{journal}{Eur. Phys. J. C}
  \textbf{\bibinfo{volume}{72}}, \bibinfo{pages}{2168} (\bibinfo{year}{2012}),
  \eprint{1206.5160}.

\bibitem[{\citenamefont{Soyez}(2019)}]{Soyez:2018opl}
\bibinfo{author}{\bibfnamefont{G.}~\bibnamefont{Soyez}},
  \bibinfo{journal}{Phys. Rept.} \textbf{\bibinfo{volume}{803}},
  \bibinfo{pages}{1} (\bibinfo{year}{2019}), \eprint{1801.09721}.

\bibitem[{\citenamefont{Aaij et~al.}(2018)}]{LHCb:2018roe}
\bibinfo{author}{\bibfnamefont{R.}~\bibnamefont{Aaij}} \bibnamefont{et~al.}
  (\bibinfo{collaboration}{LHCb}) (\bibinfo{year}{2018}), \eprint{1808.08865}.

\end{thebibliography}
%================================================================================================

%%%%%%%%%%%%%%%%%%%%%%%%%%%%%%%%%%%%%%%%%%%%%%%%%%%%%%%%%%%%%%%%
\end{document}